# COSMIC METAL PRODUCTION AND THE CONTRIBUTION OF QSO ABSORPTION SYSTEMS TO THE IONIZING BACKGROUND

*Piero Madau* [1] *and J. Michael Shull* [2,3]

[1] Space Telescope Science Institute, 3700 San Martin Drive, Baltimore MD 21218

[2] JILA, University of Colorado and National Institute of Standards and Technology, Boulder, CO 80309

[3] also at Center for Astrophysics and Space Astronomy, Department of Astrophysical, Planetary, and Atmospheric Sciences, University of Colorado, Boulder, CO 80309

## ABSTRACT

The recent discovery by Cowie et al. and Tytler et al. of metals in the Ly$\alpha$ clouds shows that the intergalactic medium (IGM) at high redshift is contaminated by the products of stars, and suggests that ionizing photons from massive star formation may be a significant contributor to the UV background radiation at early epochs. We assess the validity of the stellar photoionization hypothesis. Based on recent computations of metal yields and O-star Lyman continuum (LyC) fluxes, we find that 0.2% of the rest-mass energy of the metals produced is radiated as LyC. By modeling the transfer of ionizing radiation through the IGM and the rate of chemical enrichment, we demonstrate that the background intensity of photons at 1 Ryd that accompanies the production of metals in the Ly$\alpha$ forest clouds may be significant, approaching $0.5\times 10^{-21}$ ergs cm$^{-2}$ s$^{-1}$ Hz$^{-1}$ sr$^{-1}$ at $z\approx 3$ if the LyC escape fraction is $\gtrsim 0.25$. Together with quasars, massive stars could then, in principle, provide the hydrogen and helium LyC photons required to ionize the universe at high redshifts. We propose that observations of the He II Gunn-Peterson effect and of the metal ionization states of the Ly$\alpha$ forest and Lyman-limit absorbers should show the signature of a stellar spectrum. We also note that the stellar photoionization model fails if a large fraction of the UV radiation emitted from stars cannot escape into the IGM, as suggested by the recent *Hopkins Ultraviolet Telescope* observations by Leitherer et al. of low-redshift starburst galaxies, or if most of the metals observed at $z\approx 3$ were produced at much earlier epochs.

*Subject headings*: cosmology: observations – diffuse radiation – intergalactic medium – quasars: absorption lines

## 1. INTRODUCTION

It is widely believed that the background ultraviolet flux with energy greater than 1 Ryd arises from quasars (Sargent et al. 1980; Bergeron & Stasinska 1986; Madau 1992), possibly augmented by other sources such as hot, massive stars in star-bursting young galaxies (Bechtold et al. 1987; Steidel & Sargent 1989) or more exotic mechanisms such as decaying neutrinos (Sciama 1993). This radiation is responsible for maintaining the ionization state of the diffuse intergalactic medium, the Ly$\alpha$ forest clouds, and the metal-rich absorption systems. Using a model-independent argument from metal production, Songaila, Cowie, & Lilly (1990) suggested that the ionization state of intergalactic material could be attributed to star-forming galaxies if at least a few percent of galaxy formation occurred at redshifts $z = 3 - 5$. This possibility was studied in more detail by Miralda-Escudé & Ostriker (1990).

To date, a significant population of primeval, star-forming galaxies at high-$z$ has not been detected. However, recent measurements (Tytler et al. 1995; Cowie et al. 1995; Fan 1994) show that a large fraction of the numerous, Ly$\alpha$ forest clouds are associated with measurable C IV absorption lines. Because the Ly$\alpha$ clouds have large filling factors and may contain a significant fraction of the baryons in the universe, the new data are the first evidence for a widespread distribution of metals in the IGM. If QSO absorption systems trace the bulk of star formation occurring in galaxies at high redshifts, and if the initial mass function (IMF) extends to massive O-type stars, the stellar contribution to the local ionizing background could be significant.

In this paper, we examine the contribution of massive stars to the ionizing background, using a new calibration of the LyC emission from O stars, accurate radiative transfer of this radiation through the IGM, and a parameterization of the metal formation rate in QSO absorption-line systems. In § 2 we summarize the baryon content and metallicity of gas in Ly$\alpha$ forest clouds, Lyman-limit systems, damped Ly$\alpha$ systems, and star-forming galaxies. In § 3 we estimate the luminosity of ionizing photons that accompany the metal production and compute the spectrum of ionizing radiation. In § 4 we discuss the implications for the ionizing background and chemical evolution of the universe.

## 2. HEAVY ELEMENT ABUNDANCES AT HIGH REDSHIFT

### 2.1. Lyman-$\alpha$ Forest Clouds

The term "Ly$\alpha$ forest" is used to denote the numerous population of H I absorption systems observed in the spectrum of quasars (Weymann 1993; Bechtold 1994). With H I column densities ranging from $10^{17}$ cm$^{-2}$ down to $10^{13}$ cm$^{-2}$ and probably even lower, these systems are usually interpreted as optically thin clouds associated with the era of baryonic infall and galaxy formation. The clouds are observed to evolve rapidly with redshift between $1.7 < z < 4$ (e.g., Murdoch et al. 1986) and typically do not show metal lines in $\sim 1$ Å resolution spectra (Sargent et al. 1980; Tytler & Fan 1994; but see Lu 1991). However, recent spectra at high S/N and high resolution obtained with the HIRES spectrograph on the *Keck* telescope (Tytler et al. 1995; Cowie et al. 1995) have shown that 50%–60% of the Ly$\alpha$ forest clouds with $\log N_{\rm HI} > 14.5$ have undergone some chemical enrichment, as evidenced by weak, but measurable C IV absorption lines. The typical inferred metallicities, $Z_{\rm Ly\alpha}$, range from 0.003 to 0.01 of solar values, subject to uncertainties of photoionization models (e.g., Bergeron & Stasinska 1986; Steidel 1990; Madau 1991; Reimers & Vogel 1993) and possible contributions from local sources (Giroux, Sutherland, & Shull 1994; Viegas

& Gruenwald 1991; Gruenwald & Viegas 1993; Petitjean, Rauch, & Carswell 1994; Viegas & Friaca 1995). Clearly, these metals were produced in stars that formed in denser gas; the metal-enriched gas was then expelled from the regions of star formation into the IGM. Although carbon need not be formed by massive stars, we assume that the IMF extends as a power law to O-type stars of mass $16 - 85\ M_\odot$.

The available data do not tightly constrain the mass density parameter, $\Omega_{\text{Ly}\alpha}$, of the Ly$\alpha$ clouds. Recent spectroscopic observations of close quasar pairs (Bechtold et al. 1994; Dinshaw et al. 1994) reveal that the forest clouds have characteristic sizes much larger than previously thought, with transverse radii $\gtrsim 200 h_{50}^{-1}$ kpc, where $h_{50}$ is the Hubble constant in units of 50 km s$^{-1}$ Mpc$^{-1}$. Although the modeling is typically done for spheres, one can generalize by considering either spheres of radius $R$ or clouds with large aspect ratios, $a \equiv R/\ell = 10 a_{10}$, of transverse size $R$ to thickness $\ell$. Assuming photoionization equilibrium with the diffuse UV field and neglecting those unlikely cases in which the line of sight enters the cloud through its edge, we find that the total hydrogen gas density of a cloud of size $R = R_{200} \times 200$ kpc and observed H I column density $N_{\text{HI}} = 10^{14} N_{14}$ cm$^{-2}$ is

$$n_{\text{H}} = \left[\frac{\Gamma_{\text{HI}} N_{\text{HI}} \cos\theta}{(1+2\chi)\alpha_A(T)\ell}\right]^{1/2} \simeq (5.6 \times 10^{-5}) R_{200}^{-1/2} N_{14}^{1/2} J_{-21}^{1/2} (a\cos\theta)^{1/2}\ \text{cm}^{-3}, \qquad (1)$$

where $\Gamma_{\text{HI}}$ is the hydrogen photoionization rate, $\theta$ is the angle between the symmetry axis of the cloud and the line of sight, $\chi = n_{\text{He}}/n_{\text{H}} \approx 1/12$, $J_{912} = J_{-21}(z) \times 10^{-21}$ ergs cm$^{-2}$ s$^{-1}$ Hz$^{-1}$ sr$^{-1}$ is the metagalactic flux at the Lyman edge, $\alpha_A = 4.1 \times 10^{-13} T_4^{-0.73}$ cm$^3$ s$^{-1}$ is the (case-A) recombination rate coefficient to all levels of hydrogen, and $T = T_4 \times 10^4$ K is the cloud temperature. For an ionizing radiation field with constant energy spectral index $\alpha$, $\Gamma_{\text{HI}} \approx (1.2 \times 10^{-11}\ \text{s}^{-1}) J_{-21}/(\alpha+3)$. We adopt $T_4 = 4$ and $\alpha = 0.5$ for the background radiation. Note that setting $\langle a\cos\theta \rangle = 1$ recovers the usual limit of spherical clouds.

Press & Rybicki (1993) parameterized the observed frequency of absorbers per unit redshift by $dN/dz \simeq 4.2(1+z)^{2.46}$ for lines of rest equivalent width $> 0.32$ Å, corresponding to columns $N_{\text{HI}} \geq 1.5 \times 10^{14}$ cm$^{-2}$ ($b = 30$ km s$^{-1}$). In an Einstein-deSitter cosmology ($q_0 = 0.5$), the volume filling factor of this subpopulation is $f \approx 4.2(H_0 R/a\ c\ \cos\theta)(1+z)^{4.96}$, assuming that all clouds have similar sizes. Denoting with $n_{\text{H,crit}}$ the closure hydrogen density, we find for $\Omega_{\text{Ly}\alpha} = f n_{\text{H}}/n_{\text{H,crit}}$ at $z = 3$,

$$\Omega_{\text{Ly}\alpha} \approx 0.02\ h_{50}^{-1} \left(\frac{J_{-21} R_{200}}{a_{10}\cos\theta}\right)^{1/2}. \qquad (2)$$

We will show below that massive stars associated with QSO absorption line systems may generate an ionizing flux $J_{-21} \approx 0.5$ at these redshifts.

In order to correctly estimate the contribution of the Ly$\alpha$ forest to $\Omega_{\text{Ly}\alpha}$ over the entire range of column densities, a specific model for the dynamics of these clouds must be adopted. Values in excess of the cosmological baryon density of the universe derived from nucleosynthesis, $\Omega_b h_{50}^2 = 0.05 \pm 0.01$ (Walker et al. 1991), would be obtained in the case of spherical clouds (Meiksin & Madau 1993; Press & Rybicki 1993). We conclude that the large Ly$\alpha$ forest clouds must have large aspect ratios and contain a large fraction of the baryons in the universe.

2.2. *Lyman-Limit Systems*

The scarcer Lyman-limit systems ($\log N_{\rm HI} > 17.2$), which are optically thick to radiation with energy greater than 1 Ryd, are associated with the halo regions of bright galaxies, and represent essentially the same population of absorbers as the Mg II and C IV metal-line systems (Steidel 1993; Stengler-Larrea et al. 1995). Steidel (1990) used the column density measurements of H I and associated ions of heavy elements, in conjunction with photoionization models, to estimate the physical parameters of these systems. Although there are many uncertainties in the analysis, the typical metal abundances are $Z_{\rm LLS} \sim 0.003 Z_\odot$ at redshift $z \sim 3$, and their mass density parameter is

$$\Omega_{\rm LLS} \simeq 0.008 h_{50}^{-1} , \qquad (3)$$

comparable to the mass density of visible matter in nearby galaxies. Recent observations obtained with the *Keck* telescope have confirmed that the metallicity of a sample of Lyman-limit systems with $N_{\rm HI} \sim 3 \times 10^{17}\,{\rm cm}^{-2}$ is comparable to that of Ly$\alpha$ forest clouds with $N_{\rm HI} \gtrsim 3 \times 10^{14}\,{\rm cm}^{-2}$ (Cowie et al. 1995).

### 2.3. Damped Lyman-$\alpha$ Absorption Systems

The damped Ly$\alpha$ systems ($\log N_{\rm HI} > 20.3$) may be the progenitors of present-day luminous galaxies (Wolfe 1990). While relatively rare, they account for most of the neutral hydrogen seen at high redshifts, with a mass density parameter evolving for $0.008 < z < 3.5$ as

$$\Omega_{\rm D}(z) \simeq (3.8 \times 10^{-4}) \exp(0.83 z) h_{50}^{-1} , \qquad (4)$$

for $q_0 = 0.5$ (Lanzetta, Wolfe, & Turnshek 1995). Thus, at $z \approx 3$, we have $\Omega_D \approx 0.005 h_{50}^{-1}$. At redshift $z \approx 2$, the typical metallicity, $Z_{\rm D}$, is approximately 1/10 of solar (Pettini et al. 1994). A "closed box" cosmic chemical evolution model which matches the observed evolution of $\Omega_{\rm D}(z)$ also implies that the metallicity of the damped Ly$\alpha$ systems should decrease sharply at $z \gtrsim 3$ (Lanzetta et al. 1995).

### 2.4. Galaxies

That star-forming galaxies might rival quasars as sources of ionizing radiation is easy to demonstrate. We assume, for purposes of estimation, that each population is dominated by objects near the break of their luminosity function. (This approximation is not rigorously true, since a galaxy luminosity function with a steep slope at $L \ll L^*$ might result in dwarfs dominating the luminosity density.) Within factors of 3, the LyC photon luminosities of typical $L^*$ disk galaxies and quasars are given by $S(L^*) \approx 10^{53}$ s$^{-1}$ and $S(QSO) \approx 10^{56}$ s$^{-1}$. For example, the total LyC production from the Milky Way was recently inferred by COBE observations of [N II] 205 $\mu$m emission to be $3.5 \times 10^{53}$ s$^{-1}$ (Bennett et al. 1994), while the value quoted for QSOs is appropriate for a total luminosity $\sim 10^{46}$ ergs s$^{-1}$. If a fraction $\langle f_{\rm esc} \rangle$ of the LyC from O stars escapes the galactic H I layer into the IGM, then the ratio of the LyC production rates of $L^*$ galaxies to that of QSOs at redshifts $z = 2-4$ is given by

$$\left[\frac{\phi(L^*)}{\phi(QSO)}\right] \left[\frac{S(L^*)}{S(QSO)}\right] \langle f_{\rm esc} \rangle \approx 10 \langle f_{\rm esc} \rangle , \qquad (5)$$

where we have adopted comoving densities $\phi(L^*) = 10^{-2}$ Mpc$^{-3}$ and $\phi(QSO) = 10^{-6}$ Mpc$^{-3}$. Clearly, if $\langle f_{\rm esc} \rangle \approx 0.1$, the two sources are comparable. From models of the escape of Galactic

LyC photons from OB associations through "H II chimneys" in the H I disk layer, Dove & Shull (1994) conclude that this escape fraction is at least 14%. The actual value could be somewhat higher if one accounts for the dynamical effects of superbubbles in puncturing the H I layer. We return to this issue in § 4.

One problem with the above estimate is that we may not be justified in extrapolating the low-redshift density of bright, star-forming galaxies to $z > 3$. In fact, despite considerable observational effort, a significant population of primeval, young galaxies at high-$z$ has to date evaded detection. While the bulk of the $B \lesssim 24$ galaxies are normal dwarfs at low redshifts (Colless et al. 1993), analyses of number counts, colors, and the lack of many Lyman-break candidates suggest that more than 90% of the faint, blue galaxies with $25 < B \lesssim 27$ lie at $z < 3$ (Guhathakurta, Tyson, & Majewski 1990; Steidel & Hamilton 1993; Madau 1995). To reconcile these observations with the hypothesis that galaxies are major contributors to the ionizing background at high-$z$ requires that we invoke dwarf or low surface brightness galaxies, undetectable to $B = 27$ at $z > 3$. Such galaxies are likely to have shallow gravitational potential wells, thus allowing easy ejection of heavy elements produced during their active star-forming phase.

### 3. IONIZING PHOTON PRODUCTION

*3.1. The Ionizing Spectrum of Metal-Forming Galaxies*

If the metals detected in the Ly$\alpha$ clouds arise from a population of stars with an IMF extending to massive stars ($10 - 100\ M_\odot$), this star formation will be accompanied by the production of hot gas, heavy elements, and ionizing stellar radiation. Sutherland & Shull (1995) used the evolutionary tracks of Schaller et al. (1992), the recently computed heavy element yields for massive stars and Type II supernovae (Woosley & Weaver 1995), and a recent recalculation of O-star LyC photon production rates (Vacca, Garmany, & Shull 1995) to derive a conversion efficiency of rest mass to ionizing continuum,

$$\eta_{\rm LyC} \equiv \frac{\int_0^\infty \int_{\nu_0}^\infty L_\nu d\nu dt}{M_{\rm m} c^2} \approx 0.002 \ . \qquad (6)$$

Here, $M_{\rm m}$ is the mass of heavy elements ($Z \geq 6$) produced in massive stars and $L_\nu$ is the spectral luminosity (ergs s$^{-1}$ Hz$^{-1}$) of ionizing photons with energies above $h\nu_o = 1$ Ryd. This conversion efficiency did not change by more than 10% when the IMF slope ($\Gamma$) of stars on the upper main sequence [$dN(>M)/dM = AM^{-\Gamma}$ for $8\ M_\odot \leq M \leq 85\ M_\odot$] was varied from $\Gamma = 1.0$ to 2.0, where $\Gamma = 1.35$ corresponds to the usual Salpeter function (see also Songaila et al. 1990). The increased metal yields from high-mass stars are compensated for by a similar increase in the LyC production.

Figure 1$a$ shows the time-history of the ionizing spectrum from one such starburst. Except for the updated OB- and Wolf-Rayet stellar spectra, the resulting composite model is similar to the young-galaxy spectra of Bruzual & Charlot (1993). The stellar He II ionizing continuum above 4 Rydbergs is dominated by Wolf-Rayet stars, for which we use the model spectra of Schmutz, Leitherer, & Gruenwald (1993). For the purposes of this paper, we shall assume that the starburst duration is sufficiently short compared to a Hubble time that the relevant quantity of interest is the time-integrated spectrum (Fig. 1$b$). This should provide a reasonable estimate of the ensemble-average of a number of starburst populations.

*3.2. The Ultraviolet Background*

The mean specific intensity $J_\nu$ of the radiation background at the observed frequency $\nu_o$, as seen by an observer at redshift $z_o$ is:

$$J(\nu_o, z_o) = \frac{1}{4\pi} \int_{z_o}^{\infty} dz \frac{d\ell}{dz} \frac{(1+z_o)^3}{(1+z)^3} \epsilon(\nu, z) \exp[-\tau_{\rm eff}(\nu_o, z_o, z)] , \qquad (7)$$

where $\epsilon(\nu, z)$ is the proper volume emissivity (in ergs s$^{-1}$ cm$^{-3}$ Hz$^{-1}$) at frequency $\nu = \nu_o(1+z)/(1+z_o)$, $d\ell/dz$ is the line element in a Friedman cosmology, and $\exp[-\tau_{\rm eff}(\nu_o, z_o, z)]$ is the average transmission of a clumpy universe. We have computed the effective photoelectric optical depth $\tau_{\rm eff} \equiv -\ln(\langle e^{-\tau}\rangle)$ of a cloudy intergalactic medium following Madau (1995). For Poisson-distributed clouds we have (Paresce, McKee, & Bowyer 1980)

$$\tau_{\rm eff}(\nu_o, z_o, z) = \int_{z_o}^{z} dz' \int_0^\infty dN_{\rm HI} \frac{\partial^2 N}{\partial N_{\rm HI} \partial z'} (1 - e^{-\tau}) , \qquad (8)$$

where $\partial^2 N/\partial N_{\rm HI} \partial z'$ is the redshift and column density distribution of absorbers along the line of sight, and $\tau$ is the H I continuum optical depth through an individual cloud of column density $N_{HI}$. When $\tau \ll 1$, $\tau_{\rm eff}$ becomes equal to the mean optical depth. In the opposite limit, the obscuration is picket fence-type, and the effective optical depth becomes equal to the mean number of optically thick systems along the line of sight. The cosmic LyC opacity is dominated by absorbers with $\tau \sim 1$. Note that the He I contribution to the attenuation is negligible if QSOs contribute substantially to the ionizing background, while He II absorption[1] on the way must be included when $(1+z) > (1+z_o)(\lambda_o/228 \text{\AA})$.

The ionizing emissivity associated with metal formation can be written as

$$\epsilon(\nu, z) = \left(\frac{E_\nu}{\int_{\nu_0}^{\infty} E_\nu d\nu}\right) \eta_{\rm LyC} \, \Omega_{\rm abs} \rho_{\rm crit}(z) c^2 \dot{Z} \langle f_{\rm esc}\rangle , \qquad (9)$$

where the first term in parentheses is the normalized, time-integrated spectrum of our template star-forming galaxy (see Fig. 1$b$), $\Omega_{\rm abs}$ is the baryonic mass density parameter of metal-forming material in QSO absorption-line systems, and $\dot{Z}$ is the metal formation rate. To parameterize in simple form the many possible evolutionary models for the history of cosmic metal formation, we assume that

$$\dot{Z} = {\rm const} = \frac{Z(z_o)}{[t_{\rm H}(z_o) - t_{\rm H}(z_{\rm on})]} , \qquad (10)$$

where $t_{\rm H}$ is the Hubble time, and $z_{\rm on}$ flags the cosmic epoch at which metal formation turns on. Equations (7), (9), and (10) yield a background flux proportional to the product $\Omega_{\rm abs} Z(z_o) \langle f_{\rm esc}\rangle$. As discussed in § 2, the gaseous baryons in the IGM add up to $\Omega_{\rm abs}$ in the range $0.01 h_{50}^{-1}$ to $0.05 h_{50}^{-1}$, while the metallicities $Z$ range between 0.01 and 0.1 times the solar value of 0.02. Although the product $\Omega_{abs}Z$ may be highest in the damped Ly$\alpha$ systems, the LyC escape fraction from such systems may be lower. Note that, at high redshifts, the ionizing background will depend only weakly on the assumed metallicity evolution history. At early epochs, sufficient neutral hydrogen is contained in the IGM to significantly attenuate the LyC flux from all but the nearby sources, within $\Delta z \approx 0.2$. Because of this large optical depth, our estimate of the metagalactic flux at $z_o \approx 3$ will depend only on the "local" metal formation rate $\dot{Z}$.

---

[1] An approximate (within 5%) formula for the effective H I continuum opacity at the observed wavelength $\lambda_o \leq 912$ Å is $\tau_{\rm eff}(\lambda_o, z_o, z) \simeq 0.25 x_c^3 \, (x^{0.46} - x_o^{0.46}) + 9.4 x_c^{1.5} \, (x^{0.18} - x_o^{0.18}) - 0.7 x_c^3 (x_o^{-1.32} - x^{-1.32}) - 0.023(x^{1.68} - x_o^{1.68})$, where $x_c \equiv (\lambda_o/912 \text{\AA}) x_o$, $x_o \equiv (1+z_o)$, and $x \equiv (1+z)$ (Madau 1995). The first term on the right-hand side represents the contribution of Ly$\alpha$ clouds, the others are due to Lyman-limit systems.

Figure 2 shows the $300 - 1000\,\text{Å}$ spectrum of the metagalactic radiation field from massive stars, computed at $z_o = 3$ for different values of $z_{on} = 3.2, 3.5$, and $4.0$ and filtered through a cloudy IGM. We assume a value of $\Omega_{abs}Z(z_o)\langle f_{esc}\rangle$ equal to $1 \times 10^{-6}$, corresponding for example to $\Omega_{abs} = 0.02$ (comparable to the baryon density in the Ly$\alpha$ forest clouds), $Z = 2 \times 10^{-4}$ (about one percent of solar, the inferred metallicity of the Ly$\alpha$ forest and Lyman-limit absorbers), and $\langle f_{esc}\rangle = 0.25$. This product is quite uncertain; we return in § 4 to a discussion of values of the individual terms. In the three cases considered, the background intensity at the hydrogen Lyman edge, $J_{912}$, is found to range from 0.13 to $0.36 \times 10^{-21} h_{50}^2 \text{ ergs}\,\text{cm}^{-2}\,\text{s}^{-1}\,\text{Hz}^{-1}\,\text{sr}^{-1}$. These values are a substantial fraction of direct measurements of the metagalactic ionizing flux from the "proximity effect" (Bajtlik, Duncan, & Ostriker 1988; Bechtold 1994) and are comparable to the estimated contribution from observed QSOs (Meiksin & Madau 1993).

## 4. DISCUSSION

Recent observations of the cosmological mass density of metals associated with QSO absorption systems at high-$z$ are expected to bear significantly on the history of star formation and the chemical and ionization evolution of the universe as a whole. We have shown that the LyC photons which accompany the production of metals in the Ly$\alpha$ forest clouds could contribute significantly to the ionizing background at high redshifts. In this scenario, the metals are formed from massive stars, probably in dense gaseous regions of young galaxies, and are then expelled to large distances consistent with the sizes of Ly$\alpha$ clouds. As a consequence, it is difficult to ascertain the exact location of the LyC production except in an average sense. Similarly, the photoionization rates derived here do not distinguish between "local" sources (from a single star-forming galaxy) and the integrated effects of many such galaxies. One would naturally expect a detailed model to include fluctuations in the spectrum due to varying local star-formation rates and "filtering" by surrounding H I gas (the parameters $\dot{Z}$ and $\langle f_{esc}\rangle$ in eq. [9]).

The model calculations we just described are based on a number of parameters, some of which contain significant uncertainties. In order to provide an assessment of the validity of the stellar photoionization hypothesis, we believe that it is important to summarize the weak spots in the argument. As shown in equation (9), the LyC emissivity is proportional to the product $\Omega_{abs}Z(z_o)\langle f_{esc}\rangle$, multiplied by a spectral template and an efficiency, $\eta_{LyC} \approx 0.002$, with which LyC radiation is generated during the production of heavy elements. The spectral template is a fairly robust quantity, since it is based on the composite, time-integrated spectrum of massive stars produced in a burst of star formation. Likewise, the efficiency $\eta_{LyC}$ is nearly constant, as long as the stellar population extends as a power-law to massive stars (50–80 $M_\odot$). The metallicity of the Ly$\alpha$ clouds was based on detection of absorption by C IV, and not oxygen. Since carbon is produced primarily by low-mass stars, there is always the possibility that star formation at high redshift may have occurred with an anomalous IMF lacking massive stars. We find this possibility slim, but the detection of O VI in the Ly$\alpha$ forest clouds would settle the issue, since oxygen is unambiguously produced in massive stars (Woosley & Weaver 1995).

Next, we address the possible uncertainties associated with our estimate of the starburst contribution to the ionizing background. We choose to relate these issues to uncertainties in the three key terms in equation (9): the baryon density, $\Omega_{\rm abs}$, the metallicity, $Z(z_{\rm o})$, and the LyC escape fraction, $\langle f_{\rm esc} \rangle$. The first term is on a fairly solid basis; the baryon density associated with all types of Ly$\alpha$ absorbers at $z \approx 3$ is at least 0.01 of closure density and could approach the nucleosynthesis limit, $\Omega_b \approx (0.05 \pm 0.01) h_{50}^{-2}$. The Ly$\alpha$ forest potentially contains the bulk of these baryons, although the uncertainties due to cloud ionization and geometry are significant.

The second term, defining the cloud metallicity history, is somewhat more uncertain. Although the metallicities at a given epoch are as reliable as the ionization models used to derive them, the time history of metal production is less well defined. Current estimates for damped Ly$\alpha$ absorbers (Pettini et al. 1994) suggest that the bulk of metals formed between redshifts $z = 2 - 3$; no such estimates exist for the Ly$\alpha$ forest. For example, the metal formation rate near the observer redshift $z_{\rm o}$ could be negligible, and most of the element production may occur at earlier epochs. In this case, the background field at $z_{\rm o}$ would be quenched by the large continuum optical depth of the universe at those redshifts. More likely, galactic chemical history, like star formation history, is spread widely over time. In this case, our approximation of a constant metal formation rate may not be too bad.

The third term, $\langle f_{esc} \rangle$, parametrizing the LyC escape fraction, is probably the most uncertain of all. This factor depends on a variety of stellar, galactic, and interstellar quantities, with few observations to guide us. In a theoretical study of the structure of H II regions in vertically stratified galactic H I disk layers punctured by OB associations, Dove & Shull (1994) used distributions of H I column density (Dickey & Lockman 1990) appropriate for the solar vicinity, together with the ionizing photon luminosity function of OB associations (Kennicutt, Edgar, & Hodge 1989). They concluded that 14% of the LyC photons escaped through the H I layers into the halo in the 2.5 kpc region around the solar circle, and they speculated on how this fraction might change in other locales. For example, the escape of LyC from a large star-forming complex in a galactic nucleus would likely be higher than that from a distribution of smaller OB associations in a galactic disk. Similarly, the gas distributions and OB formation sites in dwarf or irregular galaxies probably differ from those in spiral disks. Finally, there is almost no information characterizing the structure and types of star-forming galaxies at high redshift.

A few observational tests might provide some help in constraining $\langle f_{\rm esc} \rangle$. Using spectral observations with the *Hopkins Ultraviolet Telescope*, Leitherer et al. (1995) have recently set limits of a few percent on the LyC escaping fraction in low-redshift starburst galaxies. Thus, at first sight, the stellar photoionization hypothesis may be untenable, at least at low redshift. However, even though their target galaxies were chosen to be bright in the near-ultraviolet, there is no guarantee that the LyC is not obscured by gas along a different orientation (cone effects). In addition, these constraints need to be obtained for dwarf galaxies as well. Searching for H$\alpha$ or Ly$\alpha$ emission might also constrain the escaping flux, although the quantitative sensitivity of this test is likely to be diminished by internal dust absorption (Binette et al. 1993). The interpretation of Ly$\alpha$ emission and LyC radiation escaping from galaxies is intrinsically difficult, since they rely on modeling of optical and near-UV radiation to determine the total ionizing radiation produced. We conclude, therefore, that $\langle f_{\rm esc} \rangle$ is an unconstrained parameter of our model and of previous models (Songaila et al. 1990; Miralda-Escudé & Ostriker 1990) of stellar contributions to the intergalactic radiation field.

Despite these caveats, we believe that our estimates provide some rationale for the hypothesis that hot stars could rival quasars as a source of photoionization of the IGM. If this is true, the signature of a stellar spectrum should appear in several places, notably the He II Gunn-Peterson effect, whose detected strength (Jakobsen et al. 1994) requires a relatively soft radiation field at 4 Ryd compared to 1 Ryd (Madau & Meiksin 1994; Giroux, Fardal, & Shull 1995). On the other hand, if N(He II)/N(H I) in the Ly$\alpha$ clouds is measured to be less than 100, then stars are unlikely to be major contributors to the ionizing background; after filtering by the IGM the stellar spectra are too steep. Likewise, the distribution of metal ionization states of the Ly$\alpha$ forest and Lyman-limit absorbers should reflect an ionizing stellar contribution, from either local sources or the integrated background.

Therefore, we propose three crucial tests of the stellar photoionization scenario: (1) confirming the presence of massive stars in the high-$z$ stellar IMF by detecting O VI $\lambda\lambda 1032, 1038$ absorption in the Ly$\alpha$ clouds; and (2) constraining the LyC escape fraction, $\langle f_{\rm esc} \rangle$, by far-UV continuum and Ly$\alpha$ emission observations of low-redshift starburst galaxies and by indirect estimates for star forming regions in the Galaxy; and (3) measuring the ratio N(He II)/N(H I) in the lines of sight to high-$z$ quasars.

We thank Ralph Sutherland for his help in deriving the "starburst galaxy" ionizing spectrum, and we acknowledge helpful conversations on the subject of this paper with Mark Giroux, John Stocke, and David Tytler. Support for this work was provided by the Colorado Astrophysical Theory Program (NASA grant NAGW-766). JMS thanks the Visitor Program at the Space Telescope Science Institute where this work was completed.

# REFERENCES

Bajtlik, S., Duncan, R. C., & Ostriker, J. P. 1988, ApJ, 327, 570

Bechtold, J. 1994, ApJS, 91, 1

Bechtold, J., Crotts, A. P. S., Duncan, R. C., & Fang, Y. 1994, ApJ, 437, L83

Bechtold, J., Weymann, R. J., Lin, Z., & Malkan, M. A. 1987, ApJ, 315, 180

Bennett, C. L. et al. 1994, ApJ, 434, 587

Binette, L., Wang, J. C. L., Luo, Z., & Magris, G. 1993, AJ, 105, 797

Bergeron, J., & Stasinska, G. 1986, A&A, 169, 1

Bruzual, G., & Charlot, S. 1993, ApJ, 405, 538

Colless, M., Ellis, R. S., Broadhurst, T. J., Taylor, K., & Peterson, B. A. 1993, MNRAS, 261, 19

Cowie, L. L., Songaila, A., Kim, T.-S., & Hu, E. M. 1995, AJ, 109, 1522

Dickey, J. M., & Lockman, F. J. 1990, ARA&A, 28, 215

Dinshaw, N., Impey, C. D., Foltz, C. B., Weymann, R. J., & Chaffee, F. H. 1994, ApJ, 437, L87

Dove, J. B., & Shull, J. M. 1994, ApJ, 430, 222

Fan, X.-M. 1994, Ph.D. Thesis, Columbia University

Giroux, M., Fardal, M., & Shull, J. M. 1995, ApJ, 451, in press

Giroux, M., Sutherland, R. S., & Shull, J. M. 1994, ApJ, 435, L97

Gruenwald, R., & Viegas, S. M. 1993, ApJ, 415, 534

Guhathakurta, P., Tyson, J. A., & Majewski, S. R. 1990, ApJ 357, L9

Jakobsen, P., Boksenberg, A., Deharveng, J. M., Greenfield, P., Jedrzejewski, R., & Paresce, F. 1994, Nature, 370, 35

Kennicutt, R. C., Edgar, B. K., & Hodge, P. W. 1989, ApJ, 337, 761

Lanzetta, K. M., Wolfe, A. M., & Turnshek, D. A. 1995, ApJ, 440, 435

Leitherer, C., Ferguson, H. C., Heckman, T. M., & Lowenthal, J. D. 1995, preprint

Lu, L. M. 1991, ApJ, 379, 99

Madau, P. 1991, ApJ, 376, L33

Madau, P. 1992, ApJ, 389, L1

Madau, P. 1995, ApJ, 441, 18

Madau, P., & Meiksin, A. 1994, ApJ, 433, L53

Meiksin, A., & Madau, P. 1993, ApJ, 412, 34

Miralda-Escudé, J., & Ostriker, J. P. 1990, ApJ, 350, 1

Murdoch, H. S., Hunstead, R. W., Pettini, M., & Blades, J. C. 1986, 309, 19

Paresce, F., McKee, C., & Bowyer, S. 1980, ApJ, 240, 387

Petitjean, P., Rauch, M., & Carswell, R. F. 1994, A&A, 291, 29

Pettini, M., Smith, L. J., Hunstead, R. W., & King, D. L. 1994, ApJ, 426, 79

Press, W. H., & Rybicki, G. B. 1993, ApJ, 418, 585

Reimers, D., & Vogel, S. 1993, A&A, 276, L13

Sargent, W. L. W., Young, P. J., Boksenberg, A., & Tytler, D. 1980, ApJS, 42, 41

Schaller, G., Schaerer, D., Meynet, G., & Maeder, A. 1992, A&AS, 96, 269

Schmutz, W., Leitherer, C., & Gruenwald, R. 1993, PASP, 104, 1164

Sciama D. W. 1993, Modern Cosmology and the Dark Matter Problem (Cambridge: Cambridge University Press)

Songaila, A., Cowie, L. L., & Lilly, S. J. 1990, ApJ, 348, 371

Steidel, C. C. 1990, ApJS, 74, 37

Steidel, C. C. 1993, in The Environment and Evolution of Galaxies, ed. J. M. Shull, H. A. Thronson, (Dordrecht: Kluwer), 263

Steidel, C. C. & Hamilton, D. 1993, AJ, 105, 2017

Steidel, C. C. & Sargent, W. L. W. 1989, ApJ, 343, L33

Stengler-Larrea, E. A. et al. 1995, ApJ, 444, 64

Sutherland, R. S., & Shull, J. M. 1995, in preparation

Tytler, D., & Fan, X.-M. 1994, ApJ, 424, L87

Tytler, D., Fan, X.-M., Burles, S., Cottrell, L., David, C., Kirkman, D., & Zuo, L. 1995, in QSO Absorption Lines, Proc. ESO Workshop, ed. J. Bergeron, G. Meylan, & J. Wampler (Heidelberg: Springer)

Vacca, W. D., Garmany, C. D., & Shull, J. M. 1995, ApJ, submitted

Viegas, S. M., & Friaca, A. C. S. 1995, MNRAS, 272, L35

Viegas, S. M., & Gruenwald, R. 1991, ApJ, 377, 39

Walker, T. P., Steigman, G., Schramm, D. N., Olive, K. A., & Kang, H. 1991, ApJ, 376, 51

Weymann, R. J. 1993, in The Environment and Evolution of Galaxies, ed. J. M. Shull, H. A. Thronson, (Dordrecht: Kluwer), 213

Wolfe, A. M. 1990, in The Interstellar Medium in Galaxies, ed. H. A. Thronson & J. M. Shull (Dordrecht: Kluwer), 387

Woosley, S., & Weaver, T. 1995, ApJS, in press

# FIGURE CAPTIONS

**Figure 1:** Modeled input spectrum (Sutherland & Shull 1995) from a gaussian starburst of width $2\sigma = 4$ Myr containing 5000 $M_\odot$ of stars between 8 to 85 $M_\odot$ with IMF slope $\Gamma = 1.6$ (see § 3.1). Spectrum shows four bands: soft UV (5.0 – 13.6 eV) and ionizing continua of H I (13.6 eV), He I (24.58 eV), and He II (54.4 eV). (a) Time history of ionizing spectral luminosity at times 2, 4, 6, 12, and 16 Myr (peak is at 6 Myr). (b) Time-integrated spectrum ($E_\nu$ in eq. [9]) of the starburst normalized to the total radiated ionizing energy $\int_{\nu_0}^{\infty} E_\nu d\nu$, integrated over the 42.5 Myr lifetime of an 8 $M_\odot$ star.

**Figure 2:** Observed spectrum at redshift $z_o = 3$ of a UV background dominated by massive stars, modified by the absorption of intervening Ly$\alpha$ clouds and Lyman-limit systems. We assume a constant metal-formation rate from $z_{on}$ to $z_o$ and adopt $\Omega_{abs} Z(z_o) \langle f_{esc} \rangle = 1 \times 10^{-6}$ (see § 3.2). The metagalactic flux is plotted in units of $(h_{50}^2)$ ergs cm$^{-2}$ s$^{-1}$ Hz$^{-1}$ sr$^{-1}$. From top to bottom, the three curves correspond to different epochs, $z_{on} = 3.2, 3.6$, and $4.0$, for the onset of heavy element formation. The spectrum for $z_{on} = 3.2$ shows bumpy structure because the stellar input spectrum undergoes less redshift smearing.

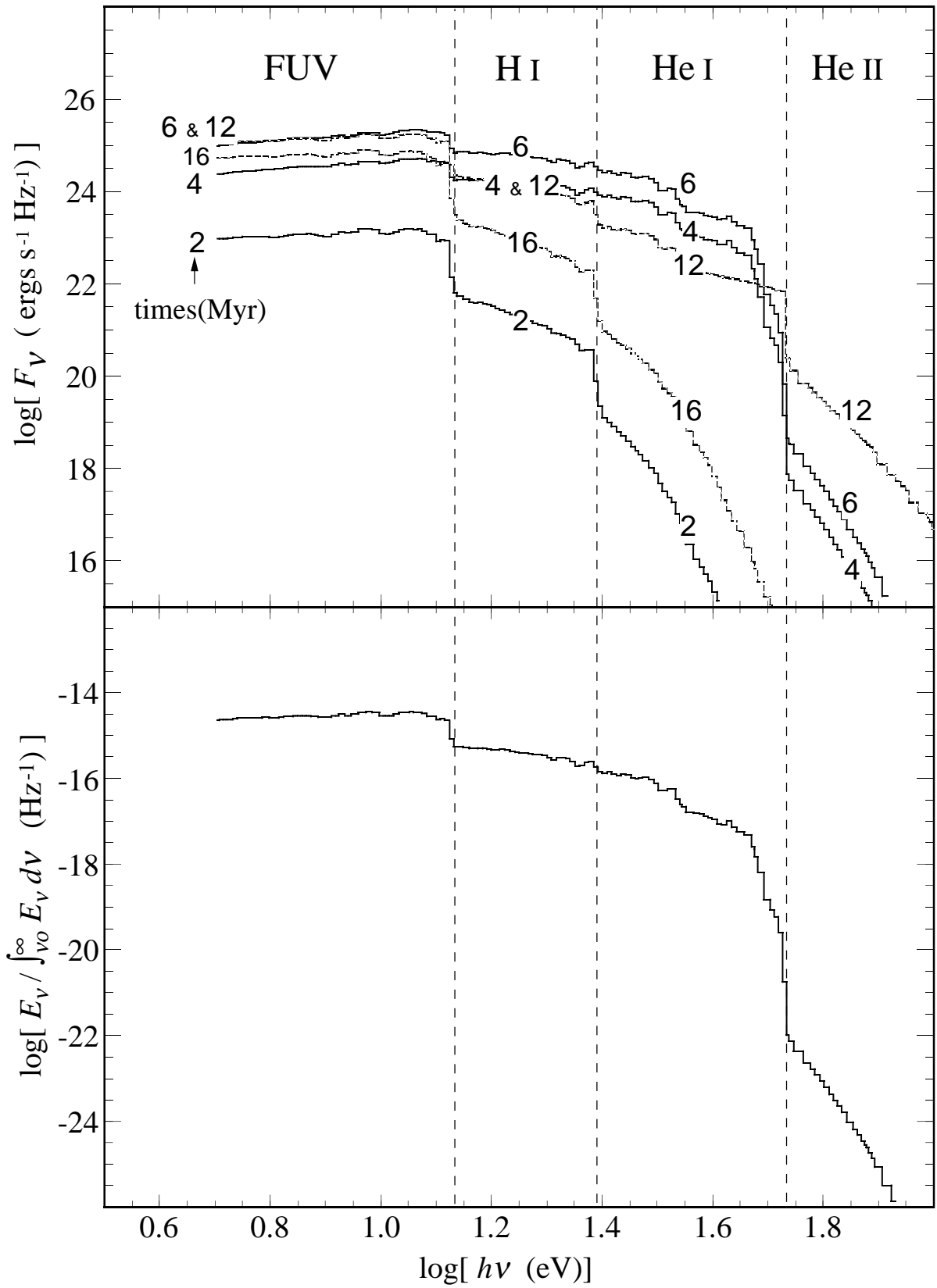

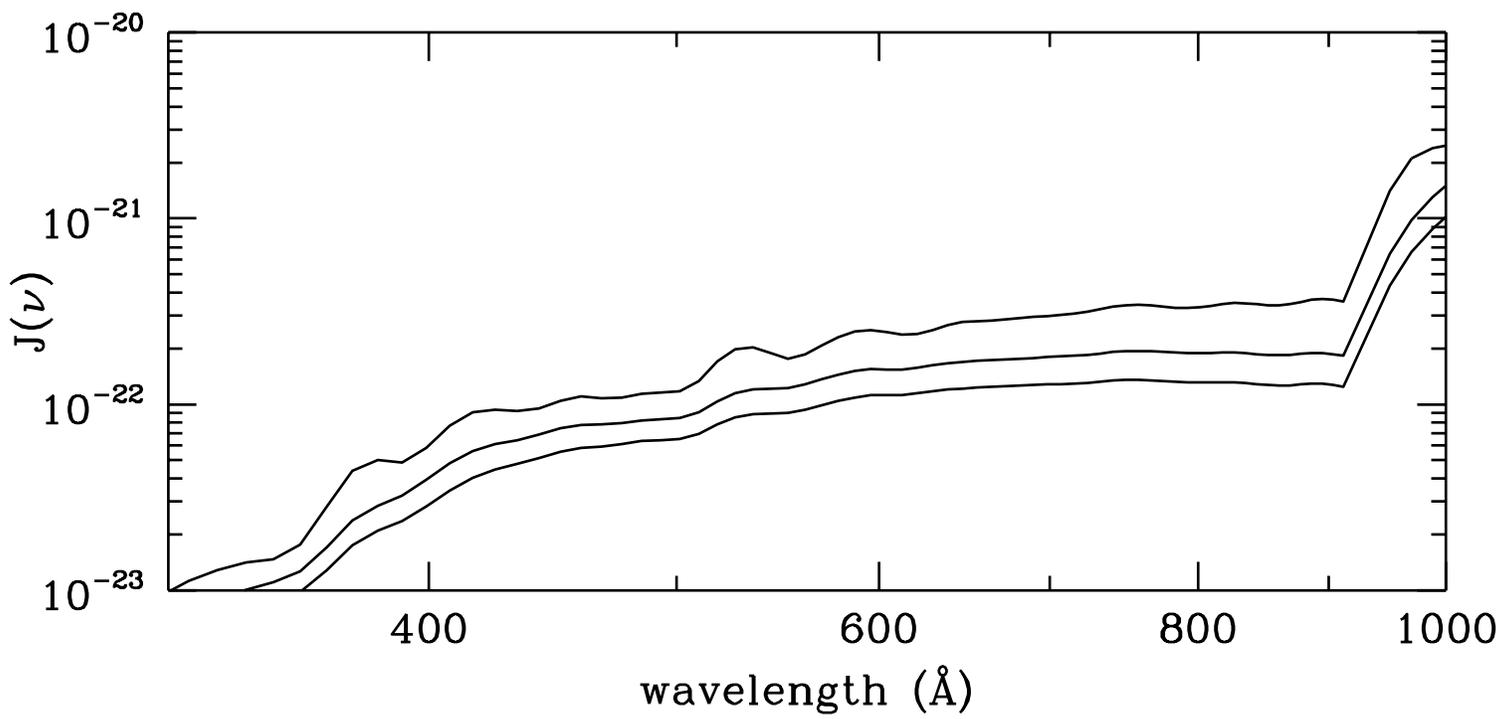